\documentclass[11pt,a4paper]{article}
\usepackage{amsfonts,graphicx}  %latexsym,graphicx,

\newcommand{\1}{\mathbb{I}}

\newcommand{\G}{\Gamma_n}
\newcommand{\cG}{-\!\!\!\Gamma_n}

\newlength{\xunit}
\newlength{\yunit}

\title{Two particles on a star graph I}
\author{Mark Harmer\\
email: harmer@ujf.cas.cz\\
Czech Academy of Sciences\\
\v{R}e\v{z}\\
Czech Republic}

\begin{document}

\maketitle 

\begin{abstract}
We consider a two particle system on a star graph with $\delta$-function interaction. A class of eigensolutions is described which are constructed from appropriate one particle solutions, and hence are parametrised by two momenta. These solutions include a family of solutions with discontinuous derivative on the diagonal.
\end{abstract}

\section{Introduction}
The bulk of research in quantum graphs has been in single particle systems. The author is aware of only one paper \cite{Mel:Pav} in which a two particle system, viz. two particle scattering on a Y graph with inter particle interaction localised at the vertex, is considered on a quantum graph. \\
Here we consider two particles, now on a star graph with $n$ edges, with $\delta$-function inter particle interaction. In particular the interaction occurs along a codimension one subset so that, unlike in \cite{Mel:Pav}, the system is an infinite rank perturbation of the `non-interacting' system. Consequently, it is not possible to use extension theory to produce explicit solutions to the problem from the non-interacting problem. \\
Part of the motivation for studying this problem is that the equivalent problem on the line ($n=2$) is completely solvable, not just for two but for an arbitrary number of particles (the literature here is extensive: we refer only to the seminal paper of Yang \cite{Yan} along with more recent work \cite{Alb:Kur, AFK1, AFK2} from the point of view of extension theory). It is natural to ask how the inclusion of a vertex affects this solvability. \\
We begin by describing all possible one-particle solutions, with the appropriate boundary conditions at the vertex, on the configuration space cut along the support of the $\delta$-function interaction. The set of all solutions to the two-particle problem, constructed using these new one-particle solutions, yields a family of eigensolutions with discontinuous derivatives on the diagonal. Moreover, a large family of solutions---not only the anti-symmetrisation of one particle solutions---is inherited from the one particle problem. We show that no more solutions may be constructed in this way. We conclude with a brief discussion of generalisation to more than two particles.
\section{Description of the two particle problem}
In this paper we consider a two particle system on a graph with $n$ semi-infinite edges conected at a single node, $\G$. The configuration space for two particles on $\G$ is $\G^2 = \G \times \G$.
\begin{figure}[ht]\hspace*{-45mm}
%\begin{center}
\includegraphics{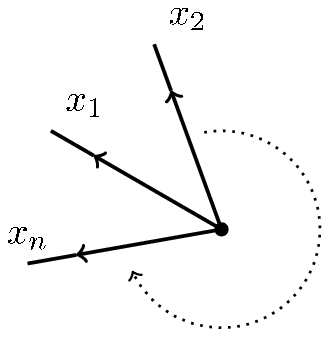}\vspace*{-230mm}
\caption{Non-compact star graph with $n$ edges, $\G$.}\label{Gn}
%\end{center}
\end{figure}
The local coordinate neighbourhoods consist of $n^2$ quadrants $(x_i, y_j)\in Q_{ij} = [0,\infty)\times[0,\infty)$, $i,j\in\{1,\ldots ,n\}$, with certain obvious identifications between the boundaries of the quadrants. \\
The dynamics on $\G^2$ is defined by the Schr\"{o}dinger type operator on each quadrant $Q_{ij}$
\begin{equation}\label{Hc}
\left. H_c \right|_{Q_{ij}} = \mbox{} - \frac{\partial^2}{\partial^2 x_i} - \frac{\partial^2}{\partial^2 y_j} + c\, \delta_{ij}\, \delta(x_i - y_i)
\end{equation}
with $c$ real. Neglecting for now the behaviour at the boundaries of the quadrants, it is known (lemma 7.1.2 of \cite{Alb:Kur}) that $H_c$ is essentially self-adjoint on the set of functions satisfying 
\begin{equation}\label{dbc}
\left. \frac{1}{2} \left( \frac{\partial \psi}{\partial x_i} - \frac{\partial \psi}{\partial y_i} \right) \right|_{Q_{ii}, x_i = y^+_i} - \left. \frac{1}{2} \left( \frac{\partial \psi}{\partial x_i} - \frac{\partial \psi}{\partial y_i} \right) \right|_{Q_{ii}, x_i = y^-_i} = c\cdot \left. \psi \right|_{Q_{ii}, x_i = y_i} 
\end{equation}
on the diagonal $D=\left\{ x_i=y_i \, ; \, x_i,y_i\in Q_{ii}, i\in\{1,\ldots ,n \} \right\}$. We complete the definition of $H_c$ by requiring the standard Kirchhoff boundary conditions at the boundaries of the quadrants
\begin{eqnarray}
\left. \psi \right|_{Q_{ij}, x_{i}=0} = \left. \psi \right|_{Q_{kj}, x_{k}=0} \; ; \quad \sum^n_{l=1} \left. \frac{\partial \psi}{\partial x_l} \right|_{Q_{lj}, x_{l}=0} = 0 \, , \; \forall i, j, k \label{bbc1} \\
\left. \psi \right|_{Q_{ij}, y_{j}=0} = \left. \psi \right|_{Q_{ik}, y_{k}=0} \; ; \quad \sum^n_{l=1} \left. \frac{\partial \psi}{\partial y_l} \right|_{Q_{il}, y_{l}=0} = 0 \, , \; \forall i, j, k \, . \label{bbc2}
\end{eqnarray}
\section{Set of basic solutions}
We define $\cG^2$ as the complex constructed from $\G^2$ by cutting each of the diagonal quadrants $Q_{ii}$ along the lines $x_i = y_i$. 
\begin{figure}[ht]\hspace*{-45mm}
%\begin{center}
\includegraphics{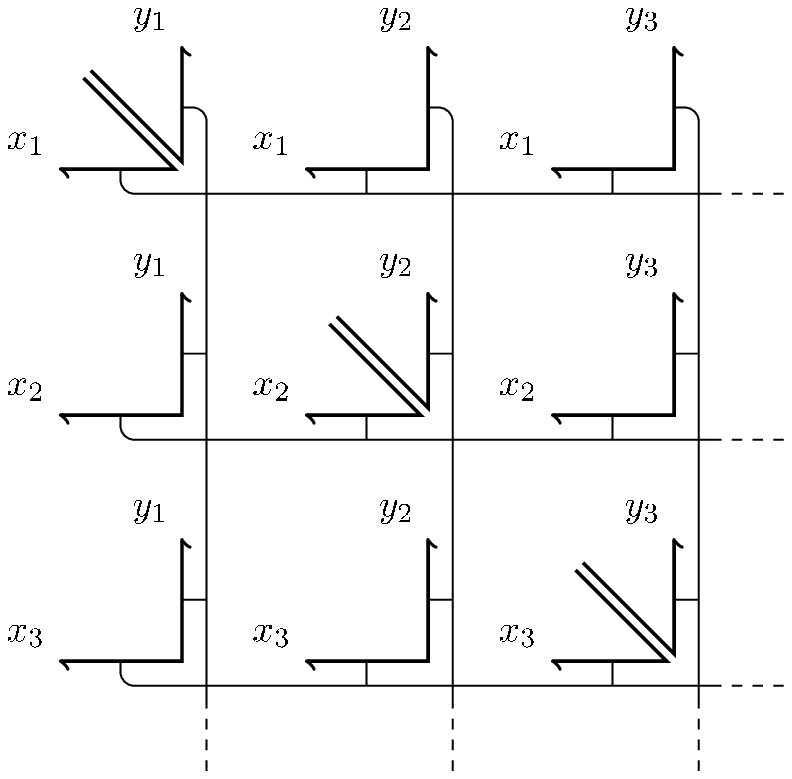}\vspace*{190mm}
\caption{The configuration space cut along the diagonal, $\cG^2$, with boundary edges identified.}\label{cG}
%\end{center}
\end{figure}
We consider all {\em one-particle} eigenstates on $\cG^2$, i.e functions of one of $x$ or $y$ which may not be smooth across the diagonal $D$, which satisfy the boundary conditions (\ref{bbc1}, \ref{bbc2}) and which are solutions of the eigenvalue equation 
$$
\mbox{} - \frac{d^2 \psi}{d^2 z} (z,k) = k^2 \psi \,.
$$
Here $z$ is one of $x_i$ or $y_j$ on $Q_{ij}$, $i,j\in\{ 1,\ldots ,n\}$. \\
It turns out that for a given particle ($x$) and momentum ($k$) there are only $n+1$ such one particle states. The first $n$ of these are inherited from the one particle states on $\G$. On the quadrant $Q_{jm}$ these may be written 
$$
\psi^l (x_j,k) = \delta^l_j\, e^{ikx_j} + S^l_j\, e^{-ikx_j} \, ,
$$
where $l,j,m \in \{ 1\ldots ,n\}$, see for example \cite{Har4}. Here $S^l_{j}$ is a unitary matrix given by $S=2P-\1$ where $P$ is the projection onto $(1,1, \ldots ,1)^t$. It will be more convenient for us to use the basis
\begin{eqnarray*}
\phi^0 & = & \frac{1}{2} \sum^n_{j=1} \psi^j \\
\phi^j & = & \frac{1}{2i} \left( \psi^j - \psi^{j+1} \right) \, ,
\end{eqnarray*}
$j\in\{1,\ldots ,n-1 \}$ ($\phi^0$ is then the eigenstate which is equal to $\cos(kx)$ everywhere on $\cG^2$ while $\phi^j$ is equal to $\sin(kx)$ on the $j$-th edge and $-\sin(kx)$ on the $j+1$-th edge of the graph). \\
The last eigenstate on $\cG^2$ 
$$
\phi^n (x_i,k) = \left\{ \begin{array}{ll}
\sin (kx_i) & : x_i\in Q_{ii} , \, x_i>y_i \;\; \mbox{or} \;\; x_i\in Q_{ij} , \, i\neq j \\ 
\left( 1 - n \right) \sin (kx_i) & : x_i\in Q_{ii} , \, x_i<y_i
\end{array} \right. 
$$
is not smooth on $D$. The one particle eigenstates for the second particle ($y$), $\{ \phi^i (y,k) \}^n_{i=0}$, are defined in the obvious way. \\
To see that this exhausts all possibilities we observe first that one particle eigenstates which are cosine on one edge are necessarily smooth across the diagonal because of (\ref{bbc1}, \ref{bbc2}). Eigenstates which are sine on one edge may be non smooth across the diagonal. Enumerating all sine solutions with discontinuity across the diagonal we see that, up to linear combinations of $\{\phi^j\}^{n-1}_{j=1}$, the `smooth sine solutions', there is only the one possible non smooth solution, $\phi^n$. \\
The two particle states constructed from the above one particle states are denoted by
$$
\Phi^{ij} \left( x, y, ; k_1, k_2 \right) = \phi^i (x,k_1)\cdot\phi^j (y,k_2) \, .
$$
These are classified according to whether they are smooth on the diagonal and whether they are symmetric or antisymmetric under the map $(k_1,k_2)\mapsto (-k_1,-k_2)$. Consequently, there are four subbases: the $(n-1)^2+1$-dimensional smooth, symmetric subbasis; the $2(n-1)$-dimensional smooth, antisymmetric subbasis; the $n-1$-dimensional non smooth, symmetric subbasis; and the 2-dimensional non smooth, antisymmetric subbasis:
\begin{equation}\label{bbas}
\left\{ \Phi^{00} \, , \, \Phi^{ij} \right\}^{n-1}_{i,j=1} \; ; \; \left\{ \Phi^{0i} \, , \, \Phi^{i0} \right\}^{n-1}_{i=1} \; ; \; \left\{ \Psi^i = \Phi^{in} - \Phi^{ni} \right\}^{n-1}_{i=1} \; ; \;  \left\{ \Phi^{0n} \, , \, \Phi^{n0} \right\}
\end{equation}
respectively. Here we should note that $\Phi^{nn}$ and $\{ \Phi^{in}+\Phi^{ni} \}^{n-1}_{i=1}$ are smooth on the diagonal $D$ (and hence can be written in terms of the smooth subbasis). 
\section{Eigensolutions of $H_c$}
We fix two unequal momenta $k_1$ and $k_2$ and consider the (augmented) smooth subbasis
\begin{equation}\label{cbas}
\left\{ \Phi^{ij}_{12} \, , \, \Phi^{ij}_{21} \right\}^{n-1}_{i,j=0}
\end{equation}
and non smooth subbasis
\begin{equation}\label{dbas}
\left\{ \Psi^{i}_{12} = \Psi^i (k_1,k_2) \, , \, \Psi^{i}_{21} = \Psi^i (k_2,k_1) \right\}^{n-1}_{i=1} \bigcup \left\{ \Phi^{0n}_{12} \, , \, \Phi^{0n}_{21} \, , \, \Phi^{n0}_{12} \, , \, \Phi^{n0}_{21} \right\}  
\end{equation}
of double the dimension. Here $\Phi^{ij}_{12} = \Phi^{ij} (k_1,k_2)$, $\Phi^{ij}_{21} = \Phi^{ij} (k_2,k_1)$. \\
In this section we describe all vectors generated by (\ref{cbas}, \ref{dbas}) which satisfy the boundary condition (\ref{dbc}) on $D$ thereby giving eigensolutions of $H_c$ with eigenvalue $k^2_1 +k^2_2$. There are three families of eigensolutions: two families constructed from the subbasis of smooth solutions and a third family with discontinuous derivative on the diagonal.
\subsection{Smooth solutions with support outside the diagonal quadrants}\label{fam1}
Using the fact that $\phi^i$, $i\in\{ 1,\ldots ,n-1\}$, has support on only two `adjacent' edges we may construct solutions with support outside the diagonal quadrants $Q_{ii}$. Such solutions trivially satisfy (\ref{dbc}).  
\begin{enumerate}
\item \mbox{}\vspace*{-4mm}
$$
\left\{ \Phi^{ij}_{12}\, , \, \Phi^{ij}_{21} \right\} \, ,
$$
for $\left\{i\in \{1,\ldots ,n-3\}, j\in \{i+2,\ldots ,n-1\} \right\}$ and \\
$\left\{i\in \{j+2,\ldots ,n-1\}, j\in \{1,\ldots ,n-3\} \right\}$. These solutions only exist for $n\ge 4$.
\item \mbox{}\vspace*{-8mm}
\begin{eqnarray*}
\left\{ \right. & \Phi^{i,i-1}_{12} + \Phi^{ii}_{12} + \Phi^{i,i+1}_{12}\, ,\,
\Phi^{i,i-1}_{21} + \Phi^{ii}_{21} + \Phi^{i,i+1}_{21}\, , & \\
& \Phi^{i-1,i}_{12} + \Phi^{ii}_{12} + \Phi^{i+1,i}_{12}\, ,\,
\Phi^{i-1,i}_{21} + \Phi^{ii}_{21} + \Phi^{i+1,i}_{21} & \left. \right\} \, ,
\end{eqnarray*}
for $i\in \{2,\ldots ,n-2\}$. These solutions only exist for $n\ge 4$.
\item \mbox{}\vspace*{-4mm}
$$
\left\{ \Phi^{12}_{12} - \Phi^{21}_{12}\, , \, \Phi^{12}_{21} - \Phi^{21}_{21} \right\} \, .
$$
These solutions exist for $n\ge 3$.
\end{enumerate}
 This gives a total of $2(n-3)(n-2) + 4(n-3) + 2 = 2n^2 - 6n + 2$ solutions.
\subsection{Smooth antisymmetric solutions}\label{fam2}
The $6n-2$ independent vectors, in the subbasis of smooth solutions (\ref{cbas}), which were not used in section \ref{fam1} all have (independent) support on the diagonal quadrants. Taking the anti-symmetrisation of these vectors will give us $3n-1$ solutions which satisfy (\ref{dbc}) on $D$:
\begin{enumerate}
\item \mbox{}\vspace*{-4mm}
$$
\left\{ \Phi^{ii}_{12} - \Phi^{ii}_{21} \right\} \, ,
$$
for $i\in \{0,\ldots ,n-1\}$.
\item \mbox{}\vspace*{-4mm}
$$
\left\{ \Phi^{0i}_{12} - \Phi^{i0}_{21}\, ,\, \Phi^{0i}_{21} - \Phi^{i0}_{12} \right\} \, ,
$$
for $i\in \{1,\ldots ,n-1\}$.
\item \mbox{}\vspace*{-4mm}
$$
\left\{ \Phi^{12}_{12} + \Phi^{21}_{12} - \Phi^{12}_{21} - \Phi^{21}_{21} \right\} \, .
$$
\end{enumerate}
There are $n+2(n-1)+1=3n-1$ antisymmetric solutions (this is only true for $n\ge3$; for $n=2$ there are of course $4=n^2$ antisymmetric solutions).
\subsection{Non Smooth solutions}\label{fam3}
The above solutions are trivial in the sense that they are independent of the value of $c$. The third family of solutions, however, has discontinuous derivative on $D$. This family of $n-1$ solutions is
\begin{equation}\label{dsoln}
\Psi^{i}_{12} - \Psi^{i}_{21} + \frac{n k_1}{c} \left( \Phi^{0i}_{12} + \Phi^{i0}_{21} \right) - \frac{n k_2}{c} \left( \Phi^{0i}_{21} + \Phi^{i0}_{12} \right)
\end{equation}
for $i\in\{ 1,\ldots ,n-1 \}$. \\
There are a total of $2n^2-2n$ solutions (for $n\ge 3$).
\subsection{Non-existence of further solutions}
The enumeration of all solutions amounts to linear algebra. However, even in the simpest case, $n=3$, it is a formidable problem (24 equations in 32 unknowns). The complete description of all possible solutions can be given in the following way. \\
We first note that the solutions described in sections \ref{fam1} and \ref{fam2} clearly exhaust all solutions constructed from the subbasis of smooth solutions. \\
To describe all non smooth solutions we first define the defect of vectors which are continuous on $D$ to be the difference between $1/c$ times the jump in the derivative on $D$ and the value on $D$. Then, enumerating all vectors generated by the non smooth subbasis which are continuous on $D$, we compare their defects to the defects generated by the smooth subbasis. If the ranges of these sets of defects coincide we can construct a solution which has discontinuous derivative on the diagonal (and in this way we see that the solutions described in section \ref{fam3} are the only possible non smooth solutions). \\ 
In a little more detail: the only vectors from the subbasis of non smooth solutions (\ref{dbas}) which are continuous across $D$ are
\begin{eqnarray}
& \Psi^{i}_{12} - \Psi^{i}_{21} & \label{cs2} \\
& \Phi^{n0}_{12} + \Phi^{0n}_{21} & \label{cs4} \\
& \Phi^{n0}_{21} + \Phi^{0n}_{12} & \, . \label{cs6} 
\end{eqnarray}
The vectors (\ref{cs2}) are accounted for in section \ref{fam1}. Letting $t_i=\frac{x_i+y_i}{2}$ be the coordinate on $D$ we see that the defects of (\ref{cs4}, \ref{cs6}) are
\begin{eqnarray*}
\left. \mbox{Def} \left( \Phi^{n0}_{12} + \Phi^{0n}_{21} \right) \right|_{x_i =y_i} & = & \frac{2 k_1}{c} \cos k_1 t_i \, \cos k_2 t_i + \frac{2 k_2}{c} \sin k_1 t_i \, \sin k_2 t_i \\
& & \mbox{} + 2 \left( \frac{2}{n} - 1 \right) \sin k_1 t_i \, \cos k_2 t_i \\
\left. \mbox{Def} \left( \Phi^{n0}_{21} + \Phi^{0n}_{12} \right) \right|_{x_i =y_i} & = & \frac{2 k_2}{c} \cos k_1 t_i \, \cos k_2 t_i + \frac{2 k_1}{c} \sin k_1 t_i \, \sin k_2 t_i \\
& & \mbox{} + 2 \left( \frac{2}{n} - 1 \right) \sin k_2 t_i \, \cos k_1 t_i   
\end{eqnarray*}
It is simple to see that the range of these defects have null intersection with the range of defects of the smooth solutions---consequently, no further solutions exist.
\section{Conclusion}
In this paper we have described a class of eigensolutions, for the two particle problem on a star graph with $\delta$ interaction, which can be written as finite sums of products of certain one particle solutions. The full space of eigensolutions of $H_c$ at a given energy is infinite dimensional so that the above finite set of solutions clearly do not describe all eigensolutions. Nevertheless, by integrating over a circle $(k_1 , k_2) = k(\cos(\theta),\sin(\theta))$ in the momentum plane we generate an appropriately large set of eigensolutions and in a future publication \cite{Har10} we plan to show how these fit into the space of all eigensolutions of $H_c$. \\
An obvious generalisation of the above is to consider $p\ge 3$ particles. Here we have in mind $p$ particles on $\G$ with only {\em pairwise} $\delta$ interaction (analogous to the hamiltonian considered for instance in \cite{Yan}). The above procedure for enumerating all solutions extends to $p$ particles (and will continue to work as long as we are only considering finite sums of products of one particle solutions). We briefly describe this extension for $p=3$ (which has, in this context, all of the complication of arbitrary $p$). \\
We define as above $\cG^3$: the configuration space for three particles cut along the interaction planes $x_i=y_i$, $x_i=z_i$ and $y_i=z_i$. Given a particle ($x$) and momentum ($k$) there are again $n$ one particle states, smooth on $\G^3$, which are inherited from the one particles states on $\G$. However, there are now $2=p-1$ non smooth states on $\cG^3$, analogous to $\phi^n$, which correspond to taking the discontinuity across either the $x_i=y_i$ or the $x_i=z_i$ plane. This gives a basis of $3!(n+2)^3$ three particle states, analogous to (\ref{cbas}, \ref{dbas}), which split into smooth and non smooth subbases. We enumerate all vectors in the non smooth subbasis which are continuous on $D$ and calculate the defect of these vectors on $D$---again $D$ is the codimension one subset formed by the interaction planes $x_i=y_i$, $x_i=z_i$, $y_i=z_i$ (Note: the interaction planes lie not only on the diagonal octants $Q_{iii}$ but also on the off diagonal octants where exactly two of the three particles are on the same edge). The final step is to consider the range of the defects from the (continuous vectors in the) non smooth subbasis with the range of defects from the smooth subbasis. If there is a non trivial intersection this may be used to construct solutions with discontinuous derivative across the interaction planes. \\
The above procedure gives interesting, i.e. non smooth, solutions in the case $p=2$; however, it is not clear that it will continue to do so for $p\ge 3$ (see \cite{Har10} for further discussion on how the $p\ge 3$ particle problem may be treated).
\section*{Acknowledgements}
This paper is dedicated to the memory of Vladimir Geyler. \\
The author is grateful for support from grant LC06002 of the Ministry of Education, Youth and Sport of the Czech Republic.

\end{document}